\documentclass{iopart}
\usepackage{iopams}
\usepackage{euscript,amssymb,epsfig}

\usepackage{amsfonts,latexsym}
\usepackage{euscript,amssymb}
\usepackage{epsfig}

\usepackage{graphicx}

\newcommand{\be}[1]{\begin{equation}\label{#1}}
\newcommand{\ee}{\end{equation}}
\newcommand{\ba}[1]{\begin{eqnarray}\label{#1}}
\newcommand{\ea}{\end{eqnarray}}
\newcommand{\rf}[1]{(\ref{#1})}
\newcommand{\nn}{\nonumber}

\renewcommand{\theequation}{\arabic{section}.\arabic{equation}}

\begin{document}

%%%%%
\title[Non-relativistic limit of multidimensional gravity]{Non-relativistic limit of multidimensional gravity: exact solutions and applications}
%%%%

\author{Maxim Eingorn\footnote[1]{e-mail: maxim.eingorn@gmail.com}, Alexander Zhuk
\footnote[2]{e-mail: ai$\_$zhuk2@rambler.ru}}

\address{Astronomical Observatory and Department of
Theoretical Physics, Odessa National University, Street Dvoryanskaya 2, Odessa 65082, Ukraine}

%\date{30 September 2009}
%
%%%%%%%%%%%%%%%%%%%%%%%%%%%%%%%%%%%%%%%%%%%%%%%%%%%%%%%%%
%
%

%
\begin{abstract}
It is found the exact solution of the Poisson equation for the multidimensional space with topology
$M_{3+d}=\mathbb{R}^3\times T^d$.
This solution describes smooth transition from the newtonian behavior $1/r_3$ for distances bigger than
periods of tori (the extra dimension sizes) to multidimensional behavior $1/r^{1+d}_{3+d}$ in opposite limit.
In the case of one extra dimension $d=1$, the gravitational potential is expressed via compact and elegant formula.
These exact solutions are applied to some practical problems to get the gravitational potentials
for considered configurations. Found potentials are used to calculate the acceleration for point masses and gravitational self-energy.
%It is shown that the corrections to the effective four-dimensional gravitational constant in the Cavendish-type experiment can be within
%the measurement accuracy of Newton's gravitational constant $G_N$.
It is proposed models where the test masses
are smeared over some (or all) extra dimensions. In 10-dimensional spacetime with 3 smeared extra dimensions, it
is shown that the  size of 3 rest extra dimensions can be enlarged up to submillimeter for the case of 1TeV fundamental
Planck scale $M_{Pl(10)}$. In the models where all extra dimensions are smeared, the gravitational potential exactly
coincides with the newtonian one regardless of size of the extra dimensions.  Nevertheless, the hierarchy problem can be solved in these models.
\end{abstract}
%

%\submitto{\CQG}

\pacs{04.50.-h, 11.25.Mj, 98.80.-k}

\maketitle

%\vspace{.5cm}

%\today \\
%\mbox{}\\
%%%%%%%%%%%%%%%%%%%%%%%%%%%%%%%%%%%%%%%%%%%%%%%%%%%%%%%%%%%%%
\section{Introduction}
\setcounter{equation}{0}

%\vspace{1cm}

There are two well-known problems which are related to each other. They are the discrepancies in gravitational constant experimental data and the hierarchy problem.
Discrepancies (see e.g. Figure 2 in the "CODATA Recommended Values of the Fundamental Constants: 2006") are usually explained by extreme weakness of gravity. It is very
difficult to measure the Newton's gravitational constant $G_N$. Certainly, for this reason geometry of an experimental setup can effect on data. However, it may well be
that, the discrepancies can also be explained (at least partly) by underlying fundamental theory.  Formulas for an effective gravitational constant following from such
theory can be sensitive to the geometry of experiments.  For example, if correction to the Newton's gravitational potential has the form of Yukawa potential, then the
force due to this potential at a given minimum separation per unit test-body mass is least for two spheres and greatest for two planes (see e.g. \cite{ISL}). Therefore,
an effective gravitational constant obtained from these formulas acquires different values for different experimental setup.

The hierarchy problem - the huge gap between the electroweak scale $M_{EW}\sim 10^3$GeV and the Planck scale $M_{Pl(4)} =1.2\times 10^{19}$GeV - can be
also reformulated in the following manner: why is gravity
so weak? The smallness of $G_N$ is the result of relation $G_N = M_{Pl(4)}^{-2}$ and huge value of $M_{Pl(4)}$. The natural explanation was proposed in \cite{large,ADD}:
the gravity is strong: $G_{\mathcal{D}} = M_{Pl(\mathcal{D})}^{-(2+d)}\sim M_{EW}^{-(2+d)}$ and it happens in ($\mathcal{D}=4+d$)-dimensional spacetime.
It becomes weak when gravity is "smeared" over large extra dimensions: $G_N \sim G_{\mathcal{D}}/V_d$ where $V_d$ is a volume of internal space.

To shed light on both of these problems from new standpoint we intend to investigate multidimensional gravity in non-relativistic limit. To do that, first, we are going
to obtain a solution of $(D=3+d)$-dimensional Poisson equation in the case of toroidal extra dimensions. {}From school years we know that newtonian gravitational
potential of a body with mass $m$ has the form of $\varphi(r_3) = -G_N m /r_3$ where $G_N$ is Newton's gravitational constant and $r_3 = |{\bf r}_3|$ is magnitude of a
radius vector in three-dimensional space. This expression is the solution of three-dimensional Poisson equation with  a point-mass source (and corresponding boundary
condition $\varphi(r_3)\to 0$ for $r_3 \to +\infty$) or it can be derived from Gauss's flux theorem in three-dimensional space.
%However, it is possible that spacetime might have a dimensionality of more than four and posses a rather complex topology.
%String theory and its supersymmetric generalizations - superstring and supergravity widely use this concept and give it
%a new foundation. Therefore, i
To investigate effects of extra dimensions, it is necessary to generalize the Newton's
formula to the case of extra dimensions. Clearly, the result depends on topology of investigated models.
We consider models where $D$-dimensional spatial part of factorizable geometry is defined on a product manifold
$M_D=\mathbb{R}^3\times T^{d}$.
$\mathbb{R}^3$ describes three-dimensional flat external (our) space and $T^{d}$ is a torus
which corresponds to $d$-dimensional internal space. Let $V_{d}$ be a volume of the internal space and $a\sim V^{1/d}_{d}$
is a characteristic size of extra dimensions. Then, Gauss's flux theorem leads to the following asymptotes for
gravitational potential (see e.g. \cite{ADD} and our Appendix for alternative derivation): $\varphi \sim 1/r_3$ for $r_3 >> a$ and $\varphi \sim 1/r^{1+d}_{3+d}$
for  $r_{3+d} << a$ where $r_3$ and $r_{3+d}$ are magnitudes of radius vectors in three-dimensional and
$(3+d)$-dimensional spaces, respectively. Obviously, an exact solution of $D$-dimensional Poisson equation should show smooth transition between both of these asymptotes.
This formula gives possibility to investigate characteristic features of multidimensional gravity in non-relativistic limit which enable us to reveal extra dimensions
(or to establish experimental limitations on extra dimensions).

In our paper we obtain such exact solution for arbitrary number of the extra dimensions. In the case of one extra dimension, this expression acquires very compact and
elegant form. Found exact solution is applied to a number of practical problems, e.g. to calculate gravitational force between two spherical shells (or balls). It gives
a possibility to calculate an effective four-dimensional gravitational constant for given configurations. For example, in the case of two balls we show that, in
principle, the inverse square law experiments enable us  to detect a deviation from the Newton's gravitational constant. Then, we generalize our model to the case of
smeared extra dimensions. It means that we suppose that test bodies are uniformly smeared/spreaded over all or part of extra dimensions. We prove that the gravitational
potential does not feel the smeared extra dimensions. In particular, if all extra dimensions are smeared, then the inverse square law experiment does not show any
deviation from the ordinary Newton's formula, and this conclusion does not depend on sizes of the smeared extra dimensions. Nevertheless, the hierarchy problem can be
solved in these models.

 The paper is structured as follows. In section 2 we get the exact solution of multidimensional Poisson
 equation in the case of spacial topology $\mathbb{R}^3\times T^{d}$. This formula is applied to some practical
 problems in section 3 to get gravitational potential, gravitational acceleration of a point mass and gravitational self-energy
 for these problems. In section 4 we investigate gravitational interaction of two spherical shells. Then, in section 5 we generalize
 our model to the case of smeared extra dimensions. Here, we prove that gravitational potential does not "feel" the smeared extra dimensions
 and demonstrate that the hierarchy problem can be solved in this case.
A brief discussion of the obtained results is presented in the concluding section 6. In Appendix A, we get the expressions for the gravitational force law at small and
big separation between point masses on multidimensional manifold with topology $\mathbb{R}^3\times T^{d}$.
%%%%%%%%%%%%%%%%%%%%%%%%%%%%%%%%%%%%%%%%%%%%%%%%%%%%%%%%%%%%%%%%%%%%%%%%%%%%

\section{Multidimensional gravitational potentials}
\setcounter{equation}{0}

In $D$-dimensional space, the  Poisson equation reads
%%%%
\be{2.1}
\triangle_D\varphi_D=S_DG_{\mathcal{D}}\rho_D({\bf r}_D)\, ,
\ee
%%%%%
where $S_D=2\pi^{D/2}/\Gamma (D/2)$ is a total solid angle (square of
$(D-1)$-dimensional sphere of a unit radius), $G_{\mathcal{D}}$ is a gravitational constant in
$(\mathcal{D}=D+1)$-dimensional spacetime
and $\rho_D({\bf r}_D)=m\delta(x_1)\delta(x_2)...\delta(x_D)$.

\subsection{Spatial topology $\mathbb{R}^D$}

In the case of topology $\mathbb{R}^D$, \rf{2.1} has the following solution:
%%%%%%%
\be{2.2}
\varphi_D({\bf r}_D)=-\frac{G_{\mathcal{D}}m}{(D-2)r_D^{D-2}}\, ,\quad D\geq 3.
\ee
%%%%%%%
This is the unique solution of \rf{2.1} which satisfies the boundary condition: $\lim\limits_{r_D\rightarrow+\infty}\varphi_D({\bf r}_D)=0$. Gravitational constant
$G_{\mathcal{D}}$ in \rf{2.1} is normalized in such a way that the strength of gravitational field (acceleration of a test body) takes the form: $-d\varphi_D / d r_D = -
G_{\mathcal{D}}m/r^{D-1}_D$.

\subsection{Spatial topology $\mathbb{R}^3\times T^{d}$, $d$-arbitrary}

If topology of space is  $\mathbb{R}^3\times T^{d}$, then it is natural to impose periodic boundary
conditions in the directions of the extra dimensions:
$\varphi_D({\bf r}_3,\xi_1,\xi_2,\ldots, \xi_i,\ldots ,\xi_{d})=
\varphi_D({\bf r}_3,\xi_1,\xi_2,\ldots, \xi_i +a_i,\ldots ,\xi_{d}), \quad i=1,\ldots ,d$, where
$a_i$ denotes a period in the direction of the extra dimension $\xi_i$. Then, Poisson equation has solution
(cf. also with \cite{ADD,CB}):
%%%%%%
\ba{2.3}
&{}&\varphi_D({\bf r}_3,\xi_1,...,\xi_{d})=-\frac{G_N m}{r_3}\nn \\
&\times&\sum\limits_{k_1=-\infty}^{+\infty}...\sum\limits_{k_{d}=-\infty}^{+\infty}
\exp\left[-2\pi\left(\sum\limits_{i=1}^{d}\left(\frac{k_i}{a_i}\right)^2\right)^{1/2}r_3\right]\nn \\
&\times&\cos\left(\frac{2\pi k_1}{a_1}\xi_1\right)...\cos\left(\frac{2\pi k_{d}}{a_{d}}\xi_{d}\right)\, . \ea
%%%%%%%
To get this result we, first, use the formula
$\delta(\xi_i)=\frac{1}{a_i}\sum_{k=-\infty}^{+\infty}\cos\left(\frac{2\pi k}{a_i}\xi_i\right)$ and, second,
put the following relation between gravitational constants in four- and $\mathcal{D}$-dimensional spacetimes:
%%%%%
\be{2.4}
\frac{S_D}{S_3}\cdot\frac{G_{\mathcal{D}}}{\prod_{i=1}^{d}a_i}=G_N\, .
\ee
%%%%%
The letter relation provides correct limit when all $a_i \to 0$. In this limit zero modes $k_i=0$ give the main contribution and we obtain $\varphi_D({\bf
r}_3,\xi_1,...,\xi_{d})\rightarrow-G_N m/r_3$. \rf{2.4} was widely used in the concept of large extra dimensions which gives possibility to solve the hierarchy problem
\cite{large,ADD}. It is also convenient to rewrite \rf{2.4} via fundamental Planck scales:
%%%%%%%
\be{2.5}
\frac{S_D}{S_3}\cdot M_{Pl(4)}^{2} = M_{Pl(\mathcal{D})}^{2+d}\prod_{i=1}^{d}a_i\, ,
\ee
%%%%%%%
where $M_{Pl(4)}= G_N^{-1/2} =1.2\times 10^{19}$GeV and $ M_{Pl(\mathcal{D})}\equiv G_{\mathcal{D}}^{-1/(2+d)}$ are fundamental Planck scales in four and $\mathcal{D}$
spacetime dimensions, respectively.

In opposite limit when all $a_i \to +\infty$ the sums in \rf{2.3} can be replaced by integrals. Using the standard integrals (e.g. from \cite{PBM}) and relation
\rf{2.4}, we can easily show  that, for example, in particular cases $d=1,2$ we get desire result: $\varphi_D({\bf r}_3,\xi_1,\ldots ,\xi_d)\rightarrow-G_{\mathcal{D}}
m/[(D-2)\; r_{3+d}^{1+d}]$.

{}From \rf{2.3}, it follows that potential energy of gravitational interaction between two point masses $m^{(a)}$ and $m^{(b)}$ with radius vectors ${\bf r}_D^{(a)}$ and
${\bf r}_D^{(b)}$ reads
%%%%
\ba{abc}
&{}&U_D\left({\bf r}_D^{(a)},{\bf r}_D^{(b)}\right)=-\frac{G_Nm^{(a)}m^{(b)}}{|{\bf r}_3^{(a)}-{\bf r}_3^{(b)}|}\nn \\
&\times&\sum\limits_{k_1=-\infty}^{+\infty}...\sum\limits_{k_{D-3}=-\infty}^{+\infty}
\exp\left[-2\pi\left(\sum\limits_{i=1}^{D-3}\left(\frac{k_i}{a_i}\right)^2\right)^{1/2}|{\bf r}_3^{(a)}-{\bf r}_3^{(b)}|\right]\nn \\
&\times&\cos\left(\frac{2\pi k_1}{a_1}|\xi_1^{(a)}-\xi_1^{(b)}|\right)...\cos\left(\frac{2\pi k_{D-3}}{a_{D-3}}|\xi_{D-3}^{(a)}-\xi_{D-3}^{(b)}|\right)\, . \ea
%%%%%%

\subsection{Spatial topology $\mathbb{R}^3\times T^{1}$}

In the case of one extra dimension $d=1$ we can perform summation of series in \rf{2.3}. To do it, we can apply the Abel-Plana formula or simply use the tables of series
\cite{PBM}. As a result, we arrive at compact and nice expression:
%%%%%
\be{2.6}
\varphi_4({\bf r}_3,\xi)=-\frac{G_N m}{r_3}\frac{\sinh\left(\frac{2\pi r_3}{a}\right)}{\cosh\left(\frac{2\pi
r_3}{a}\right)-\cos\left(\frac{2\pi\xi}{a}\right)}\, ,
\ee
%%%%%
where $r_3 \in [0,+\infty )$ and $\xi \in [0,a]$. It is not difficult to verify that this formula has correct asymptotes when $r_3>>a$ and $r_4<<a$. Figure 1
demonstrates the shape of this potential. Dimensionless variables $\eta_1\equiv r_3/a \in [0,+\infty )$ and $\eta_2\equiv \xi/a \in [0,1]$. With respect to variable
$\eta_2$, this potential has two minima at $\eta_2=0,1$ and one maximum at $\eta_2 =1/2$. We continue the graph to negative values of $\eta_2 \in [-1,1]$ to show in more
detail the form of minimum at $\eta_2=0$. The potential \rf{2.6} is finite for any value of $r_3$ if $\xi\neq 0,a$ and goes to $-\infty $ as $ -1/r_4^2$ if
simultaneously $r_3 \to 0$ and $\xi \to 0,a$ (see Figure 2). We would like to mention that in particular case $\xi=0$ formula \rf{2.6} was also found in
\cite{Barvinsky}.
%%%%%%%%%%%%%%%%%%%%%
\begin{figure}[hbt]
\centerline{\epsfxsize=7cm \epsfbox{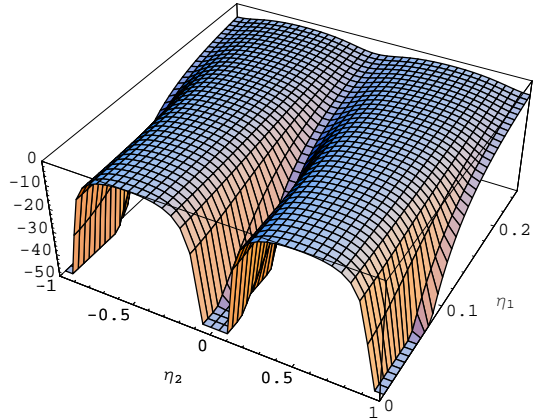} } \caption{
Graph of function $\tilde \varphi (\eta_1,\eta_2) \equiv \varphi_4({\bf r}_3,\xi)/(G_N m/a)=
-\sinh (2\pi \eta_1)/[\eta_1(\cosh(2\pi\eta_1)-\cos(2\pi\eta_2))]$. \label{potential}}
\end{figure}
%%%%%%%%%%%%%%%%%%%%%%%%%%%%

%%%%%%%%%%%%%%%%%%%%%
\begin{figure}[hbt]
%\centerline{
\centerline{\epsfxsize=7cm \epsfbox{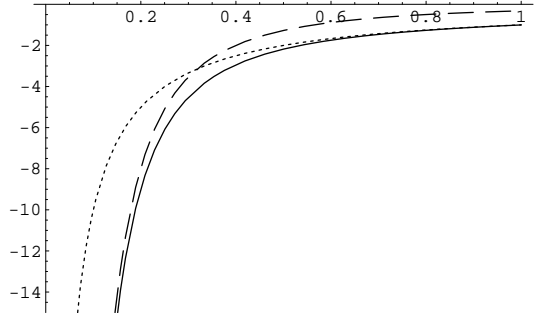} } \caption{
Section $\xi =0$ of potential \rf{2.6}. Solid line is $\tilde \varphi (\eta_1,0) =
-\sinh (2\pi \eta_1)/[\eta_1(\cosh(2\pi\eta_1)-1)]$ which goes to $-1/\eta_1$ (dotted line) for $\eta_1 \rightarrow +\infty$
and to $-1/(\pi \eta_1^2)$ (dashed line) for $\eta_1 \rightarrow 0$. \label{asympt}}
\end{figure}
%%%%%%%%%%%%%%%%%%%%%%%%%%%%
%%%%%%%%%%%%%%%%%%%%%%%%%%%%%%%%%%%%%%%%%%%%%%%%%%%%%%%%%%%%%%%%%%%%%%%%%%%%%%%%
%%%%%%%%%%%%%%%%%%%%%%%%%%%%%%%%%%%%%%%%%%%%%%%%%%%%%%%%%%%%%%%%%%%%%%%%%%%%%%%%%
%%%%%%%%%%%%%%%%%%%%%%%%%%%%%%%%%%%%%%%%%%%%%%%%%%%%%%%%%%%%%%%%%%%%%%%%%%%%%%%%%

\subsection{Yukawa approximation}

Having at hand formulas \rf{2.3} and \rf{2.6}, we can apply it for calculation of some elementary physical problems and compare
obtained results with known newtonian expressions.
For a working approximation, it is usually sufficient to summarize in \rf{2.3} up to the first Kaluza-Klein modes $|k_i|=1\, (i=1,\ldots ,d)$:
%%%%%%
\be{2.7} \fl \varphi_D({\bf r}_3,\xi_1,\xi_2,...,\xi_d)\approx-\frac{G_N m}{r_3}\left[1 +
2\sum\limits_{i=1}^{d}\exp\left(-\frac{2\pi}{a_i}r_3\right)\cos\left(\frac{2\pi}{a_i}\xi_i\right)\right]\, . \ee
%%%%%%%
Then, the terms with the biggest periods $a_i$ give the main contributions.  If all test bodies
are on the same brane ($\xi_i=0$) we obtain:
%%%%%%%
\be{2.8} \fl \varphi_D({\bf r}_3,\xi_1=0,\xi_2=0,...,\xi_{d}=0) \equiv \varphi_D (r_3)\approx -\frac{G_N m}{r_3} \left[1+\alpha
\exp\left(-\frac{r_3}{\lambda}\right)\right]\, , \ee
%%%%%%
where $\alpha =2s \; (1 \le s\le d)\, $, $\lambda =a/(2\pi)$ and $s$ is a number of extra dimensions with periods of tori $a_i$ which are equal (or approximately equal)
to $a=\max a_i$. If $a_1=a_2= \ldots = a_d=a$, then $s=d$. Thus, the correction to Newton's potential has the form of Yukawa potential. It is now customary to interpret
tests of gravitational inverse-square law (ISL) as setting limits on additional Yukawa contribution. The overall diagram of the experimental constraints can be found in
\cite{new} (see Figure 6) and we shall use these data for limitation $a$ for given $\alpha$.
%%%%%%%%%%%%%%%%%%%%%%%%%%%%%%%%%%%%%%%%%%%%%%%%%%%%%%%%%%%%%%%%%%%%%%%%%%%%

\section{Application ($\xi_i=0$)}
\setcounter{equation}{0}

Now, we apply formulas \rf{2.3} and \rf{2.6} to some particular geometrical configurations. For our calculations we shall use the case of $\xi_1=\xi_2=\ldots =\xi_d =0$.
It means that test bodies have the same coordinates in the extra dimension. It takes place e.g. when test bodies are on the same brane. Also, to get  numerical results
we should define the sizes $a_i$ of the extra dimensions.
%If the standard model fields are not localized on the brane,
The $e^+e^-$ leptonic interaction experiments at high-energy colliders show that there is no deviations from Coulomb's law for separations down to $10^{-16}$cm
\cite{ISL}. Therefore, if the standard model fields are not localized on the brane, this value can be used for upper bound of $a_i: \quad a_i\lesssim 10^{-17}$cm (in
section 5 we demonstrate how to avoid this argument).
%give the upper bound $a_i\lesssim 10^{-17}$cm. This limitation originates from
%\footnote{Coulomb's law has been tested for separations down to $10^{-16}$cm in
%$e^+e^-$ leptonic interactions at high-energy colliders \cite{ISL}. Nevertheless, see also the footnote \rf{2} below.}.
%For this value of $a$ the 5-dimensional fundamental Planck scale is
%$M_{Pl(5)}\gtrsim 10^{11}$TeV (see \rf{5}).
These values $a_i$ can be greatly increased if we suppose that the standard model fields are localized on the brane. In this case, we can obtain the upper bound for
$a=\max a_i$ from the gravitational inverse-square law experimental data depicted in Figure 6 of paper \cite{new}. For example, the Yukawa approximation \rf{2.8} shows
that $\alpha =2$ for $d=1$. For this value $\alpha$, Figure 6 gives $\lambda \leq 4.7\times 10^{-3}{\mbox{cm}}\;\Rightarrow\; a\leq 3.0\times 10^{-2}$cm.

\subsection{Infinitesimally thin shell}

Let us consider first an infinitesimally thin shell of mass $m=4\pi R^2 \sigma$, where $R$ and $\sigma$ are  radius and surface density of
the shell. Then, gravitational potential of this shell in a point with radius vector $\bf r_3$ (from the center of the shell) is
%%%%%
\ba{3.1}
&{}&\varphi_D(r_3)=-\frac{2\pi G_N\sigma R}{r}\int\limits_{r_3-R}^{r_3+R}\sum_k\exp[-2\pi\chi_kr']dr'\\
&=&-\frac{G_N m}{r_3} - \frac{2G_N\sigma R}{r_3}\mathop{{\sum}'}_k\frac{1}{\chi_k}e^{-2\pi\chi_kr_3}\sinh(2\pi\chi_k R)\; , \quad r_3>R\nn
\ea
and
%%%%%%
\ba{3.2}
&{}&\varphi_D(r_3)=-\frac{2\pi G_N\sigma R}{r_3}\int\limits_{R-r_3}^{R+r_3}\sum_k\exp[-2\pi\chi_kr']dr'\\
&=& -\frac{G_N m}{R} - \frac{2G_N\sigma R}{r_3}\mathop{{\sum}'}_k\frac{1}{\chi_k}e^{-2\pi\chi_k R}\sinh(2\pi\chi_k r_3)\; , \quad r_3<R\nn, \ea
%%%%%%
where
%%%%%
\be{3.3}
\chi_k \equiv \left[\sum_{i=1}^{d}\left(\frac{k_i}{a_i}\right)^2\right]^{1/2}\, , \quad
\sum_k \equiv \sum\limits_{k_1=-\infty}^{+\infty}...\sum\limits_{k_{d}=-\infty}^{+\infty}
\ee
%%%%%%%%
and the prime in sums denotes that the zero mode $k_1=\ldots =k_d =0$ is absent in summation. It means that the Newton's expressions (which correspond to the zero mode)
are singled out from sums. In the case $d=1$, these expressions can  be written in the compact form:
%%%%%%
\be{3.4} \varphi_4(r_3)=-\frac{G_N\sigma Ra}{r_3}\ln\left\{\frac{\cosh\left(\frac{2\pi(r_3+R)}{a}\right)-1}{\cosh\left(\frac{2\pi(r_3-R)}{a}\right)-1}\right\}\; , \quad
r_3\gtrless R. \ee
%%%%%%
%where $\varphi(r_3)\equiv \varphi_4({\bf{r_3}},0)$.
These formulas demonstrate two features of the considered models. Firstly, we see that inside $(r_3<R)$ of the shell  gravitational potential is not a constant. Thus, a
test body undergoes an acceleration in contrast to the newtonian case (and to Birkhoff's theorem of general relativity in four-dimensional spacetime which states that
the metric inside an empty spherical cavity in the center of a spherically symmetric system is the Minkowski metric). Secondly, if $r_3 \to R$, these potentials have a
logarithmic divergency of the type: $\sum_{k=1}^{+\infty}1/k$. For example, in this limit $r_3 \to R$ \rf{3.4} has the following asymptotic behavior:
%%%%%%
\be{3.5} \fl \varphi_4(r_3)\approx -\frac{G_N m}{R}\left[1-\frac{a}{2\pi R}\ln\left(\frac{2\pi|R-r_3|}{a}\right)\right] \equiv -\frac{G_N m}{R}\left[1+\delta\right]\, ,
\ee
%%%%%%%
where we took into account $R>>a$ and $|R-r_3|<<a$. In particular case $2\pi R =10$cm and $2\pi |R-r_3| = 10^{-1}a$, the deviation $\delta$ constitutes $2.3\times
10^{-3}$ and $2.3\times 10^{-18}$ parts of the newtonian value $-G_Nm/R$ for $a = 10^{-2}$cm and $a=10^{-17}$cm, respectively. In principle, the former estimate is not
very small. However, it is very difficult to set an experiment which satisfies the condition $|R-r_3|<<a$.
%It is hardly possible to observe such deviations in experiments.
If the shell has a finite thickness, then the divergence disappears.

\subsection{Spherical shell}

Here, we consider a spherical shell of inner radius $R_1$ and outer radius $R_2$ and mass $m=4\pi\int^{R_2}_{R_1}\rho (R)R^2dR$ where $\rho (R)$ is a volume density of the shell (in the
case of constant volume density $\rho = m /\left[\frac{4\pi}{3}(R_2^3-R_1^3)\right]$). Then, the potential outside of the shell is
%%%%%%%%%%%
\ba{3.6} &{}&\varphi_D(r_3)=-\frac{2G_N}{r_3}\left\{\sum\limits_k\frac{1}{\chi_k}e^{-2\pi\chi_kr_3}\int\limits_{R_1}^{R_2}\rho(R)R\sinh(2\pi\chi_k R)dR\right\}\nn
\\ &\equiv&\varphi^{(1)}_D(r_3)\; , \quad r_3>R_2\, , \ea
%%%%%%%%%%%
which for the constant $\rho$ reads
%%%%%%
\be{3.7}
\varphi_D(r_3)=-\frac{G_N m}{r_3}-\frac{G_N\rho}{2\pi^2r_3}\mathop{{\sum}'}_k \frac{1}{\chi_k^3}\exp[-2\pi\chi_kr_3]h_k(R)|_{R_1}^{R_2}\; ,
\ee
%%%%%%%
where
%%%%
\be{3.8}
h_k(R)=2\pi\chi_kR\cosh[2\pi\chi_kR]-\sinh[2\pi\chi_kR]\, .
\ee
%%%%%%
Inside of the shell we obtain:
%%%%%%%%%%%
\ba{3.9} &{}&\varphi_D(r_3)=-\frac{2G_N}{r_3}\left\{\sum\limits_k\frac{1}{\chi_k}\sinh(2\pi\chi_k r_3)\int\limits_{R_1}^{R_2}\rho(R)Re^{-2\pi\chi_kR}dR\right\}\nn \\
&\equiv&\varphi^{(2)}_D(r_3)\; , \quad r_3<R_1\, \ea
%%%%%%%%%%%
and for the constant $\rho$:
%%%%%
\ba{3.10} &{}&\varphi_D(r_3)=-2\pi G_N\rho\left(R_2^2-R_1^2\right)\nn \\ &+&\frac{G_N\rho}{\pi r_3}\mathop{{\sum}'}_k\frac{1}{\chi_k^2}\sinh\left(2\pi \chi_k r_3\right)
\left[\left(R+\frac{1}{2\pi\chi_k}\right)e^{-2\pi\chi_kR}\right]_{R_1}^{R_2}\, . \ea
%%%%%%%
To get potential within the shell, we can use the following relation:
%%%%%%
\be{3.11} \fl \varphi_D(r_3)=\varphi^{(1)}_D(r_3|R_2=r_3)+\varphi^{(2)}_D(r_3|R_1=r_3)\; , \quad R_1\leq r_3\leq R_2. \ee
%%%%%%
Thus, in the case of constant $\rho$ the gravitational potential within the shell reads
%%%%%%
$$
\varphi_D(r_3)= 2\pi G_N\rho \left(\frac{r_3^2}{3}-R_2^2+\frac{2R_1^3}{3r_3}\right) - \frac{G_N\rho}{2\pi^2r_3}\mathop{{\sum}'}_k\frac{1}{\chi_k^3}
$$
\be{3.12} \fl\times\left[ 2\pi\chi_kr_3-\sinh(2\pi\chi_kr_3)\left(2\pi\chi_kR_2+1\right)e^{-2\pi\chi_kR_2} -e^{-2\pi\chi_kr_3}h_k(R_1)\right]\, . \ee
%%%%%%
For one extra dimension $d=1$ we obtain a compact expression which is valid for full range of variable
$r_3\geq 0$:
%%%%%%
\be{3.13}
\varphi_4(r_3)=-\frac{G_N\rho a}{r_3}\int\limits_{R_1}^{R_2}
R\ln\left\{\frac{\cosh\left(\frac{2\pi(R+r_3)}{a}\right)-1}{\cosh\left(\frac{2\pi(R-r_3)}{a}\right)-1}\right\}dR\, ,
\ee
%%%%%%
where $\rho$ is taken to be constant.

It can be easily seen that all these potentials are finite in the limit $r_3 \to R_1,R_2$ and $R_1\neq R_2$, i.e. divergency of the potential is absent for finite thickness
of the shell. However, divergency takes place for acceleration of a test body.  We can see it from exact formulas for acceleration outside of the shell:
%%%%%
$$
-\frac{d\varphi_D}{dr_3}=-\frac{G_N m}{r_3^2}
$$
\be{3.14}\fl-\frac{G_N\rho}{2\pi^2r_3^2}\mathop{{\sum}'}_k\frac{1}{\chi_k^3} \left(2\pi\chi_kr_3+1\right)\, e^{-2\pi\chi_kr_3}\, [h_k(R_2)-h_k(R_1)]<0\, , \quad r_3>R_2,
\ee
%%%%%%
within the shell:
%%%%
\ba{3.15} &{}&-\frac{d\varphi_D}{dr_3}=-\frac43 \pi G_N\rho \frac{r_3^3-R_1^3}{r_3^2}\nn \\
&-&\frac{G_N\rho}{2\pi^2r_3^2}\mathop{{\sum}'}_k\frac{1}{\chi_k^3} \left\{ h_k(r_3)\left(2\pi\chi_kR_2+1\right)e^{-2\pi\chi_kR_2}\right.\nn \\ &-&\left.
\left(2\pi\chi_kr_3+1\right)e^{-2\pi\chi_kr_3}h_k(R_1) \right\}\, ,\quad R_1<r_3<R_2 \ea
%%%%%%
and inside of the shell:
%%%%%
\be{3.16} \fl -\frac{d\varphi_D}{dr_3}=-\frac{G_N\rho}{\pi r_3^2}\mathop{{\sum}'}_k\frac{1}{\chi_k^2}h_k(r_3)
\left[\left(R+\frac{1}{2\pi\chi_k}\right)e^{-2\pi\chi_kR}\right]_{R_1}^{R_2}\geqslant 0\, , \quad r_3<R_1. \ee
%%%%%
All of these equations \rf{3.14}-\rf{3.16} have logarithmic divergency in the limit $r_3\to R_1,R_2$, e.g. in $d=1$ case:
%%%%
\be{3.17} -\frac{d\varphi_4}{dr_3} \; \longrightarrow\; \mp 2G_N \rho a \ln \frac{2\pi|R_{1,2}-r_3|}{a}\, , \ee
%%%%
where sign "-" corresponds to $r_3 \to R_1$ and sign "+" corresponds to $r_3\to R_2$.

Putting the limit $R_1\equiv 0$, $R_2\equiv R$ in expressions obtained above in this subsection, we can get the corresponding equations for a sphere. For example, using
the same conditions as for \rf{3.5}, we can get for a sphere $(d=1)$:
%%%%%%
\be{3.18} \fl -\frac{d\varphi_4}{dr_3}\approx -\frac{G_N m}{R^2}\left[1-\frac{3a}{2\pi R}\ln\left(\frac{2\pi|R-r_3|}{a}\right)\right] \equiv -\frac{G_N
m}{R}\left[1+\overline{\delta}\right].\, \ee
%%%%%%%
Therefore, deviation from the newtonian acceleration $\overline{\delta}=3\delta$ and we can conclude that this deviation
is also difficult to observe
at experiments for considered parameters.

\begin{figure}[hbt]
%\centerline{
\centerline{\epsfxsize=7cm \epsfbox{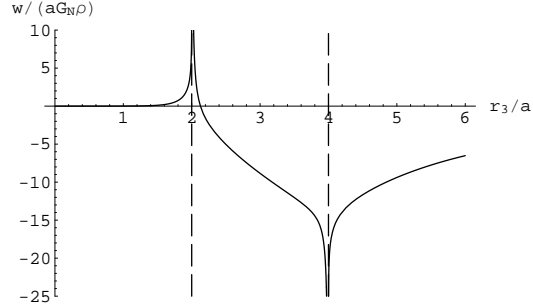} } \caption{ Graph of acceleration $w=-d\varphi_4/dr_3$ in dimensionless units (see \rf{3.14}-\rf{3.16}). Here, $a=2.5, R_1=5$
and $R_2=10$. The dashed lines correspond to radii of the shell. The rightmost line goes to the newtonian asymptote $-1/r_3^2$ when $r_3\to +\infty$. Point $r_3=0$
corresponds to unstable equilibrium.\label{acceleration}}
\end{figure}

\rf{3.14} and \rf{3.16} show that acceleration changes the sign from negative outside of the shell to positive inside of the shell (see Figure \ref{acceleration}). This
change happens within the shell.
%%%%%%%%%%%%%%%%%%%%%
%%%%%%%%%%%%%%%%%%%%%%%%%%%%

\subsection{Gravitational self-energy}

Gravitational self-energy of the spherical shell with potential $\varphi_D(r_3)$ given by formulas \rf{3.12} and \rf{3.13} reads
%%%%%%
\ba{3.19} \fl &{}& U=2\pi\rho \int_{R_1}^{R_2}r_3^2\varphi_D(r_3)dr_3 =
\frac{3G_N m^2}{10\left(R_2^3-R_1^3\right)^2}\left(-2R_2^5-3R_1^5+5R_1^3R_2^2\right)\nn \\
\fl &-&\frac{9G_N m^2}{2\left(R_2^3-R_1^3\right)^2}\mathop{{\sum}'}_k\frac{1}{(2\pi\chi_k)^2}\left[\frac{1}{3}(R_2^3-R_1^3)+\frac{1}{(2\pi\chi_k)^3}E_k\right]\, , \ea
%%%%%%
where
%%%%%%
\ba{3.20} &{}&E_k = (1+2\pi\chi_kR_2)e^{-2\pi\chi_kR_2}\left[-h_k(R_2)+2h_k(R_1)\right]\nn \\ &-&(1+2\pi\chi_kR_1)e^{-2\pi\chi_kR_1}\, h_k(R_1)\, . \ea
%%%%%

In the case of sphere $R_1=0, R_2\equiv R$ and this expression is reduced to the following formula:
%%%%%%%%
\ba{3.21} &{}&U=-\frac{3G_N m^2}{5R}\left\{1+ \frac{15}{2(2\pi R)^3}\right.\nn \\
&\times&\left.\mathop{{\sum}'}_k\frac{1}{\chi_k^3}\left[\frac{2\pi \chi_kR}{3}
-\left(\frac{1}{2\pi \chi_k R}\right)^2(1+2\pi\chi_kR)e^{-2\pi\chi_kR}h_k(R)\right]\right\}\nn \\
&\equiv& U_N+\delta U\, .
\ea
%%%%%%%%
Here, $U_N$ and  $\delta U$ are the newtonian self-energy of the sphere and deviation from it, respectively. In the case of one extra-dimension $d=1$ and for condition
$a\ll R$ we obtain:
%%%%%%
\be{3.22}
\delta U \approx U_N \frac{5}{4\pi^2}\left(\frac{a}{R}\right)^2\sum_{k=1}^{+\infty}\frac{1}{k^2} = U_N\frac{5}{24}\left(\frac{a}{R}\right)^2\, ,
\ee
%%%%%%
where we used that $\sum_{k=1}^{+\infty}1/k^2=\pi^2/6$. It is worth of noting that the correction to the newtonian formula has power law suppression instead of
exponential one as it is usually expected in Kaluza-Klein models. Nevertheless, for astrophysical objects this correction term is negligibly small because of big
difference between $a$ and $R$. For example, if $a=10^{-2}$cm, we get that $\delta U \approx 4\times 10^{-27}U_N$ and $\delta U \approx 4\times 10^{-19}U_N$ for Sun
($R\approx 7\times 10^{10}$cm) and neutron star ($R\approx 7\times 10^6$cm), respectively.
%%%%%%%%%%%%%%%%%%%%%%%%%%%%%%%%%%%%%%%%%%%%%%%%%%%%%%%%%%%%%%%%%%%%%%%%%%%%%%%%%%%%%%%%%%%%%
%%%%%%%%%%%%%%%%%%%%%%%%%%%%%%%%%%%%%%%%%%%%%%%%%%%%%%%%%%%%%%%%%%%%%%%%%%%%%%%%%%%%%%%%%%%%%

\section{Gravitational interaction of two spherical shells}
\setcounter{equation}{0}

Let us consider two spherical shell with radiuses $R_1<R_2$ and constant density $\rho$ for first shell and radiuses $R_1'<R_2'$ and constant
density $\rho'$ for second shell.
Then, potential energy of gravitational interaction between these shells reads
%%%%%%
\be{4.1}
U(r_3)= \int\int\int \varphi_D (r_3')\rho' dV'\, .
\ee
%%%%%%
Here, $r_3\geq R_2+R_2'$  is magnitude of three-dimensional vector between centers of shells and $r_3'$ is magnitude of three-dimensional vector between center of the
first shell and an arbitrary elementary mass $dm'=\rho'dV'$ within second shell. Potential $\varphi_D(r_3')$ is given by Eq. \rf{3.7}. After integration, we obtain:
%%%%%%%%
\ba{4.2} &{}& U(r_3)=-\frac{G_N m m'}{r_3} -\frac{G_N\rho\rho'}{4\pi^4 r_3}\nn
\\ &\times&\mathop{{\sum}'}_k\frac{1}{\chi_k^6}e^{-2\pi\chi_kr_3}[h_k(R_2)-h_k(R_1)][h_k(R'_2)-h_k(R'_1)]\, . \ea
%%%%%%%%
In the case of two spheres ($R_1\equiv 0, R_2\equiv R$ and $R_1'\equiv 0, R_2'\equiv R'$), this formula is reduced to:
%%%%%%%
\be{4.3}
U(r_3)=-\frac{G_N m m'}{r_3} -\frac{G_N\rho\rho'}{4\pi^4
r_3}\mathop{{\sum}'}_k\frac{1}{\chi_k^6}e^{-2\pi\chi_kr_3}h_k(R)h_k(R')\, .
\ee
%%%%%%%

\subsection{Yukawa approximation }

For working approximation, it is often sufficient to keep in sums the first Kaluza-Klein modes, i.e. to use the Yukawa approximation. In this approximation, \rf{4.2} and
\rf{4.3} can be rewritten correspondingly:
%%%%%%%%%%%%%%%%%%%%%%
\be{4.4}
U(r_3)\approx-\frac{G_N
mm'}{r_3}-\frac{16\pi^2G_N \rho\rho'\alpha\lambda^6}{r_3}\; e^{-r_3/\lambda}\; \widetilde h(R)|^{R_2}_{R_1}\; \widetilde h(R)|^{R_2'}_{R_1'}
\ee
%%%%%%%%%%%%
and
%%%%%%%
\be{4.5}
U(r_3)\approx-\frac{G_N mm'}{r_3}\left[1+9\alpha \left(\frac{\lambda}{R}\right)^3\left(\frac{\lambda}{R'}\right)^3e^{-r_3/\lambda}\right]
\widetilde h(R)\; \widetilde h(R')\, ,
\ee
%%%%%%
where
%%%%%
\be{4.6} \widetilde h(R)= \frac{R}{\lambda}\cosh\left(\frac{R}{\lambda}\right)-\sinh\left(\frac{R}{\lambda}\right). \ee
%%%%%%%%%%%%%%%%%%%%
Here, $\lambda=a/(2\pi )$ and parameter $\alpha=2s$ is described in \rf{2.8}.

\subsection{Gravitational force between two spheres}

In Yukawa approximation, gravitational force between two spheres is
%%%%%
\ba{4.7} \fl F(r_3) = -\frac{dU}{dr_3}&\approx&-\frac{G_N mm'}{r_3^2}\left[1+9\alpha \left(\frac{\lambda}{R}\right)^3\left(\frac{\lambda}{R'}\right)^3
\frac{r_3}{\lambda}\; e^{-r_3/\lambda}\right]
\widetilde h(R)\; \widetilde h(R')\nn \\
\fl &\approx& -\frac{G_N mm'}{r_3^2}\left[1+\frac{9}{4}\alpha \left(\frac{\lambda}{R}\right)^2\left(\frac{\lambda}{R'}\right)^2 \frac{r_3}{\lambda}\;
e^{-(r_3-R-R')/\lambda}\right]\, , \ea
%%%%%
where in the last expression we use conditions $R,R'\gg\lambda$. If surfaces of the spheres are on the distances of the order of the maximal period: $r_3-R-R'\sim a$,
then we get
%%%%%%
\be{4.8} \fl F(r_3) \approx -\frac{G_N mm'}{r_3^2}\left[1 + 0.0084 s \left(\frac{\lambda}{R}\right)^2\left(\frac{\lambda}{R'}\right)^2 \frac{r_3}{\lambda}\right]\equiv
-\frac{G_N mm'}{r_3^2}\left(1+ \widetilde\delta\right)\, . \ee
%%%%%%%%
For example, in the case of one extra dimension and $R=R'=1$cm, $\lambda =4.7\times 10^{-3}$cm, we get for deviation from the newtonian formula the estimate
 $\widetilde \delta =1.8\times 10^{-9}$. To obtain it, it is necessary to remember that for $d=1$ parameter $s=1$.
%%%%%%%%%%%%%%%%%%%%%%%%%%%%%%%%%%%%%%%%%%%%%%%%%%%%%%%%%%%%%%%%%%%%%%%%%%%%%%%%%%%%%%%%%%%%%%%%%%%%
%%%%%%%%%%%%%%%%%%%%%%%%%%%%%%%%%%%%%%%%%%%%%%%%%%%%%%%%%%%%%%%%%%%%%%%%%%%%%%%%%%%%%%%%%%%%%%%%%%%%
%%%%%%%%%%%%%%%%%%%%%%%%%%%%%%%%%%%%%%%%%%%%%%%%%%%%%%%%%%%%%%%%%%%%%%%%%%%%%%%%%%%%%%%%%%%%%%%%%%%%

\section{Smeared extra dimensions}
\setcounter{equation}{0}

Now, we get onto the asymmetrical extra dimension models (cf. with \cite{asymm}) with topology
%%%%%%%%%%%
\be{5.1}
M_D=\mathbb{R}^3\times T^{d-p}\times T^p \, , \quad p\le d\, ,
\ee
%%%%%%%%%%%
where we suppose that $(d-p)\, $ tori have the same "large" period $a$ and $p$ tori have "small" equal periods $b$. In this case, the fundamental
Planck scale relation \rf{2.5} reads
%%%%%%
\be{5.2}
\frac{S_D}{S_3}\cdot M_{Pl(4)}^{2} = M_{Pl(\mathcal{D})}^{2+d}a^{d-p}\, b^p\, .
\ee
%%%%%%
Additionally, we assume that test bodies are uniformly smeared/spreaded over small extra dimensions. Thus, test bodies have a finite thickness in small extra dimensions (thick brane approximation). For short, we shall call such small extra dimensions as "smeared" extra dimensions. If $p=d$ then all
extra dimensions are smeared.

It is not difficult to show that the gravitational potential does not feel smeared extra dimensions. We can prove this statement by three different methods. First, we can directly solve D-dimensional Poisson equation \rf{2.1} with
the periodic boundary conditions for the extra dimensions $\xi_{p+1},\ldots ,\xi_d$ and the mass density $\rho = \left(m/\prod_{i=1}^p a_i\right)
\delta({\bf r}_3)\delta(\xi_{p+1})...\delta(\xi_d)$. As a result, we obtain the unique solution $\varphi_D({\bf r}_3,\xi_{p+1},...,\xi_{d})$ of the form \rf{2.3} which does not depend
on $\xi_1, \ldots ,\xi_p$ and satisfies the limit $\lim\limits_{r_3\rightarrow+\infty}\varphi_D({\bf r}_3,\xi_{p+1},...,\xi_{d})=0 $.

Second, we can average solutions \rf{2.3} (where point mass $m$ is replaced by $m\prod_{i=1}^p (d\xi_i/a_i)$) over smeared dimensions $\xi_1, \ldots ,\xi_p$ and
take into account that $\int_0^a \cos(2\pi k\xi/a)d\xi =0,a$ for $k\neq 0$ and $k=0$, respectively. Thus, the terms with these variables disappear from \rf{2.3}. For example, direct integration of the compact
 expression \rf{2.6} results in pure newtonian formula:
%%%%%%
\ba{5.3} \fl &{}&-\frac{G_N m}{ar_3}\sinh\left(\frac{2\pi r_3}{a}\right)\int\limits_0^a\left[\cosh\left(\frac{2\pi r_3}{a}\right)
-\cos\left(\frac{2\pi\xi}{a}\right)\right]^{-1}d\xi \nn \\
\fl &=&-\frac{G_N m}{\pi r_3}\arctan\left[\left(\frac{\cosh\left(\frac{2\pi r_3}{a}\right)+1}{\cosh\left(\frac{2\pi
r_3}{a}\right)-1}\right)^{1/2}\tan\left(\frac{\pi\xi}{a}\right)\right]_0^a = -\frac{G_N m}{r_3} \, . \ea
%%%%%

Finally, from the symmetry of the model, it is clear that in the case of a test mass smeared over extra dimensions, the gravitational field vector ${\bf E}_D= -\nabla_D \varphi_D$
does not have components with respect to smeared extra dimensions. For example, if we consider the model with topology of a cylinder $S^1\times \mathbb{R}$
where $S^1$ is the smeared extra dimension and $\mathbb{R}$ is the external dimension, then ${\bf E}_2$ will be parallel to an element of the cylinder. For simplicity, let us consider
a model where all extra dimensions are smeared. Then, the Poisson equation \rf{2.1} can be rewritten in the form:
%%%%%%
\be{5.4} \nabla_D{\bf E}_D=-S_DG_{\mathcal{D}}\rho_D({\bf r}_D)=-4\pi G_Nm\delta({\bf r}_3)\, , \ee
%%%%%%
where we use relation \rf{2.4} between gravitational constants and formula $\rho_D({\bf r}_D)=m\delta({\bf r}_3)/\prod\limits_{i=1}^{d}a_i$.
After integrating both parts of this formula over multidimensional volume $V=V_3\cup V_{internal}$ we obtain:
%%%%%%
\be{5.5}
\int_V E^{\alpha_i}_{D\; ,\, \alpha_i} d^DV = -4\pi G_N m V_{internal }=\mbox{const} \, .
\ee
%%%%%%
Here, $d^DV=\varepsilon_{|\alpha_1,\ldots ,\alpha_D|}dx^{\alpha_1}\wedge\ldots \wedge dx^{\alpha_D}=
dxdydzdx^{i_1}\ldots dx^{i_{d}}$ and $\alpha_i=1,2,3,i_1,\ldots ,i_d$. Indexes $i_1,\ldots ,i_d$ enumerate extra dimensions.
Applying the Gauss theorem, we get
%%%%%%
\be{5.6}
\int_V E^{\alpha_i}_{D\; ,\, \alpha_i} d^DV = \int_{\partial V}E_D^{\alpha_i}d^{D-1}\Sigma_{\alpha_i}=4\pi R^{2}_3E_D V_{internal}\, ,
\ee
%%%%%
where
%%%%%%
\ba{5.7}
d^{D-1}\Sigma_{\alpha_i} = \varepsilon_{\alpha_i |\alpha_j\ldots \alpha_k|} &\underbrace{dx^{\alpha_j}\wedge \ldots \wedge dx^{\alpha_k}}&  \\
&\ \ \ \ \ \ \ \ D-1&\nn \ea
%%%%%
and in three-dimensional spherical coordinates $d^{D-1}\Sigma_{r_3}=R^2_3\sin \vartheta d\vartheta d\varphi dx^{i_1}\ldots dx^{i_{D-3}}$. To get \rf{5.6}, we use the
fact that $E^{r_3}_D\equiv E_D$ is the only non-zero component of the gravitational field vector. Therefore, comparing \rf{5.5} and \rf{5.6}, we obtain:
%%%%%
\be{5.8}
E_D(r_3)=-\frac{G_Nm}{r_3^2} \quad \Longrightarrow\quad  \varphi_D(r_3)=-\frac{G_Nm}{r_3}\, .
\ee
%%%%%%%

Thus, all these 3 approaches show that the gravitational potential does not feel smeared extra dimensions. It means that in the case of $p\, $ smeared extra dimensions,
the wave numbers $k_1,\ldots ,k_p$ disappear from \rf{2.3} and we should perform summation only with respect to $k_{p+1}, \ldots ,k_d$.
%%%%%%%%%%%%%%%%%%%%%%

\subsection{Effective gravitational constant}

As it follows from \rf{4.7}, in the Yukawa approximation, the gravitational force between two spheres with masses $m_1,m_2$, radiuses $R_1,R_2$ and distance $r_3$
between the centers of the spheres reads:
%%%%%%
\be{5.9} F(r_3) = -\frac{G_{N(eff)}(r_3) m_1m_2}{r_3^2}\, , \ee
%%%%%%
where
%%%%%
\ba{5.10} \fl &{}&G_{N(eff)}(r_3)\approx G_N \left\{1+\frac{9}{2}(d-p)\left(\frac{\lambda}{R_1}\right)^2\left(\frac{\lambda}{R_2}\right)^2 \frac{
r_3}{\lambda}e^{-(r_3-R_1-R_2)/\lambda}\right\}\nn \\ \fl &\equiv& G_N(1+\delta_G)\, \ea
%%%%%
is an effective gravitational constant.
Here, we took into account that in the case of $p\; $ smeared dimensions, the prefactor $\alpha$ should be replaced by $2(d-p)$.
Now, we want to evaluate the corrections $\delta_G$ to the Newton's gravitational constant and to estimate their possible influence on the experimental data.
%to compare it with the experimental uncertainties achieved in measurement of $G_N$.
As it follows from Figure 2 in the CODATA 2006, the most precise values of $G_N$ were obtained in the University
Washington and the University Z\"urich experiments \cite{U-wash,U-zur}. They are  $G_N/10^{-11}{\mbox m}^3{\mbox kg}^{-1}{\mbox s}^{-2} =
6.674215\pm 0.000092$,
%$G_N/10^{-11}{\mbox m}^3{\mbox kg}^{-1}{\mbox s}^{-2} =
and $6.674252\pm 0.000124$, respectively. Let us consider two particular examples: $(\mathcal{D}=5)$-dimensional model  with $d=1, p=0\; \rightarrow \alpha = 2$ and
$(\mathcal{D}=10)$-dimensional model  with $d=6, p=3\; \rightarrow \alpha = 6$. For these values of $\alpha$, Figure 6 in \cite{new} gives the upper limits for
$\lambda=a/(2\pi)$ correspondingly $\lambda  \approx 4.7\times 10^{-3}$cm and $\lambda  \approx 3.4\times 10^{-3}$cm. To calculate $\delta_G$, we take parameters of the
Moscow experiment \cite{Moscow}: $R_1\approx 0.087$cm for a platinum ball with the mass $m_1=59.25\times 10^{-3}$g, $R_2\approx 0.206$cm for a tungsten ball with the
mass $m_2=706\times 10^{-3}$g and $r_3=0.3773$cm. For both of these models we obtain $\delta_G \approx 8.91\times10^{-12}$ and $\delta_G \approx 1.06\times10^{-14}$,
respectively.
These values are rather far from the measurement accuracy of $G_N$ in \cite{U-wash,U-zur}. In future, if we can achieve in the
Moscow-type experiments the accuracy within these values, then, changing radii $R_{1,2}$ and distance $r_3$, we can reveal extra dimensions or obtain experimental
limitations on considered models. We should note that small changes in experimental bounds for $\lambda$ result in drastic changes of $\delta_G$. For example, if for
$\lambda$ we take the upper limits $\lambda  \approx 2\times 10^{-2}$cm and $\lambda  \approx 1.3\times 10^{-2}$cm which follow from previous experiments
(see Figire 5 in \cite{ISL} for $\alpha = 2$ and 6, respectively) then $\delta_G \approx 0.0006247$ and $\delta_G \approx 0.0000532$ and these figures are very close
to the measurement accuracy of $G_N$ in \cite{U-wash,U-zur}.

\subsection{Model: $\mathcal{D}=10$ with $d=6, p=3$}

Let us consider in more detail $(\mathcal{D}=10)$-dimensional model  with 3 smeared dimensions.
Here, we have very symmetrical with respect to a number of spacial dimensions structure: 3 our external dimensions, 3 large extra dimensions with periods $a$ and 3 small smeared extra dimensions with
periods $b$. For $b$ we put a limitation: $b\le b_{max} = 10^{-17}$cm which is usually taken for thick brane approximation. This limitation follows from the electrical inverse-square law experiments.
Although, in the next subsection we shaw that such approach can be significantly relaxed for models with smeared extra dimension, here we still use this bound.
%However, in the case of the smeared extra dimensions, the ISL is not sensitive to these dimensions (see the next subsection). Thus, it gives us a possibility to relax considerably this bound. %Nevertheless

\begin{figure}[htbp]
%\centerline{
\centerline{\epsfxsize=7cm \epsfbox{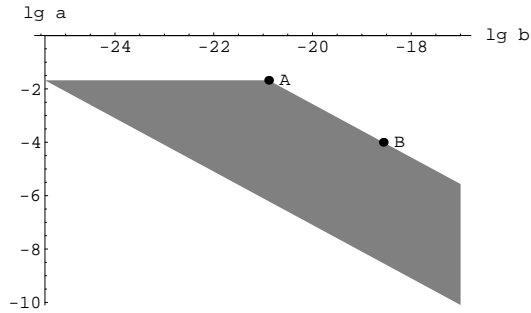} } \caption{ Allowed region (shadow area) for periods of large ($a$) and smeared ($b$) dimensions in the model
$\mathcal{D}=10$ with $d=6, p=3$. \label{trapezium}}
\end{figure}

As we mentioned above, in the case of $\alpha =6$, for $a$ we should take a limitation $a\le a_{max} =2.1\times 10^{-2}$cm.  To solve the hierarchy problem, the
multidimensional Planck scale is usually considered from 1TeV up to approximately 130 TeV (see e.g. \cite{asymm,supernova}). To make some estimates, we take
$M_{min}=1$TeV$\lesssim M_{Pl(10)}\lesssim M_{max}=50$TeV. Thus, as it follows from \rf{5.2}, the allowed values of $a$ and $b$ should satisfy inequalities:
%%%%%%%
\be{12a} \frac{S_9}{S_3}\frac{M_{Pl(4)}^2}{M_{max}^{8}}\le a^3b^3\le \frac{S_9}{S_3}\frac{M_{Pl(4)}^2}{M_{min}^{8}}. \ee
%%%%%%%
Counting all limitations, we find allowed region for $a$ and $b$ (shadow area in Figure \ref{trapezium}). In this trapezium, the upper horizontal and right vertical
lines are the decimal logarithms of $a_{max}$ and $b_{max}$, respectively. The  right and left inclined lines correspond to $M_{Pl(10)}=1$TeV and $M_{Pl(10)}=50$TeV,
respectively. To illustrate this picture, we consider two points A and B on the line $M_{Pl(10)}=1$TeV. Here, we have $a=2.1\times 10^{-2}$cm, $b=10^{-20.9}$cm for A and
$a=10^{-4}$cm, $b=10^{-18.6}$cm for B. These values of large extra dimensions $a$ are much bigger than in the standard approach $a \sim 10^{(32/6)-17}$cm $\approx
10^{-11.7}$cm \cite{large,ADD}.
%%%%%%%%%%%%%%%%%%%%
%%%%%%%%%%%%%%%%%%%%%%%%%%%%

\subsection{Model: $\mathcal{D}$-arbitrary and $d= p$}

In this model, the test masses are smeared over all extra dimensions. Therefore, in non-relativistic limit, there is no deviation from the Newton's law at all.
Surprisingly, this result does not depend on the size of smeared extra dimensions. The ISL experiments will not show any deviation from the Newton's law without regard
to the size $b$ (see also \rf{5.10} where $d-p=0$). Similar reasoning are also applicable to Coulomb's law.  It is necessary to suggest other experiments which can
reveal the multidimensionality  of our spacetime. Nevertheless, we can solve the hierarchy problem in this model because \rf{5.2} (where $d=p$) still works.  For
example, in the case of bosonic string dimension $\mathcal{D}=26$ we find $M_{Pl(26)}\approx 31$TeV for $b =10^{-17}$cm. In the case $\mathcal{D}=10$ we get
$M_{Pl(10)}\approx 30$TeV for $b =5.59\times 10^{-14}$cm. In general, if we suppose that $1 \mbox{TeV} \lesssim M_{Pl(\mathcal{D})}\lesssim 30$TeV then we obtain:
$$
p=1,\ \ \ 1.65\times 10^{11}\mbox{cm}\lesssim b\lesssim4.46\times 10^{15}\mbox{cm},\ \ \
$$
$$
p=2,\ \ \ 3.81\times 10^{-4}\mbox{cm}\lesssim b\lesssim3.43\times 10^{-1}\mbox{cm},\ \ \
$$
$$
p=3,\ \ \ 4.83\times 10^{-9}\mbox{cm}\lesssim b\lesssim1.40\times 10^{-6}\mbox{cm},\ \ \
$$
$$
\ \ \ \vdots\ \ \
$$
$$
p=6,\ \ \ 5.59\times 10^{-14}\mbox{cm}\lesssim b\lesssim5.21\times 10^{-12}\mbox{cm},\ \ \
$$
where lower limit corresponds to 30TeV and upper limit agrees with 1TeV.

\section{Conclusions}

We have considered generalization of the Newton's potential to the case of extra dimensions where multidimensional space
has topology $M_D=\mathbb{R}^3\times T^{d}$. We obtained the exact solution \rf{2.3} which describes smooth transition from the newtonian behavior $1/r_3$ for distances bigger than periods of tori (the extra dimension sizes) to multidimensional behavior $1/r^{1+d}_{3+d}$ in opposite limit.  In the case of one extra dimension,
the gravitational potential is expressed via compact and elegant formula \rf{2.6}. We applied these exact solutions to some practical problems to get the gravitational potentials
for considered configurations. Found potentials were used to calculate the acceleration for point masses and gravitational self-energy.

To estimate corrections to the newtonian expressions, it is sufficient to keep only the first Kaluza-Klein modes. Then, we obtained that the correction term has the form of the Yukawa potential with parameters defined
by multidimensional models. Such representation gave us a possibility to use the results of the inverse square-law experiments to get limitations on periods of tori.
%On basis of these limitation,
As an Yukawa potential approximation, it was shown that in the Cavendish-type experiments the corrections (due to the extra dimensions) to the newtonian gravitational constant are still
far from the measurement accuracy of experiments for the determination of $G_N$.
%It may reveal the extra dimensions or provide experimental limitations on parameters of multidimensional models.

Then, we proposed models where test masses can be smeared over the extra dimensions. The number of the smeared dimensions can be equal or less than the total number of
the extra dimensions. We proved that the gravitational potential does not feel the smeared dimensions and this conclusion does not depend on the size of these
dimensions. Such approach opens new remarkable possibilities. For example in the case $\mathcal{D}=10$ with 3 large and 3 smeared extra dimensions and $M_{Pl(10)}
=1$TeV, the large extra dimensions can be as big as the upper limit established by the ISL experiments for $\alpha =6$, i.e. $a\approx 2.1\times 10^{-2}$cm. This value
of $a$ is in many orders of magnitude bigger than the rough estimate $a \approx 10^{-11.7}$cm obtained from the fundamental Planck scale relation of the usual form
\rf{2.5}. The limiting case where all extra dimensions are smeared is another interesting example. Here, there is no deviation from the Newton's law at all. Thus, these
models can explain negative results for the detection of the extra dimensions in ISL experiments irrespective of the size of the extra dimensions. Nevertheless, these
models can be still used to solve the hierarchy problem.
%For example, in the case $\mathcal{D}=10$ (all extra dimensions are smeared) we got $M_{Pl(10)}\approx 30$TeV for $b =5.59\times 10^{-14}$cm.

%%%%%%%%%%%%%%%%%%%%%%%%%%%%%%%%%%%%%%%%%%%%%%%%%%%%%%%%%%%%%%%%%%%%%%%%%%%%%%
%%%%%%%%%%%%%%%%%%%%%%%%%%%%%%%%%%%%%%%%%%%%%%%%%%%%%%%%%%%%%%%%%%%%%%%%%%%%%%5
%\section{\label{sec:7}Summary and discussion}
%\setcounter{equation}{0}

%%%%%%%%%%%%%%%%%%%%%%%%%%%%%%%%%%%%%%%%%%%%%%%%
\ack We thank Uwe G\"{u}nther for his stimulating discussions.
A. Zh. acknowledges the hospitality
of the Theory Division of CERN and the High Energy, Cosmology and Astroparticle Physics Section of the ICTP
during preparation of this work.
This work was supported in part by the
"Cosmomicrophysics" programme of the Physics and Astronomy
Division of the National Academy of Sciences of Ukraine.
%%%%%%%%%%%%%%%%%%%%%%%%%%%%%%%%%%%%%%%%%%%%%%%%%

%\section*{Appendix:The gravitational force law under dimensional reduction}
%\setcounter{equation}{0}

\appendix
\section{\label{sec:A}The gravitational force law under dimensional reduction}
\renewcommand{\theequation}{A.\arabic{equation}}
\setcounter{equation}{0}

Let us consider a product space $M_D$ consisting of two
components $M_D=M_{3}\times M_{d}$, where for simplicity of the subsequent calculations we assume $M_{3}=\mathbb{R}^3$ as
$3-$dimensional flat external space and a $d-$torus
$M_{d}=T^{d}$ (with the same characteristic length $a$ along
each of the $d$ dimensions) as compact internal space. The
volume of $M_{d}$ is hence given as
$vol(M_{d})=V_{d}=a^{d}$.

The aim of this appendix is to demonstrate that the force laws for
the full theory on $M_D$ and the effective theory on $M_{3}$ have correspondingly the form
%%%%%%
\be{g1}
F(r_{3+d})=G_{\mathcal{D}}\frac{m_1m_2}{r_{3+d}^{2+d}}\, ,
\ee
%%%%%
%%%%%%
\be{g1a}
F(r_3)=G_{N}\frac{m_1m_2}{r_3^{2}}\, ,
\ee
%%%%%%%
where gravitational constants $G_{\mathcal{D}}$ and $G_N$ are related with each other in accordance with \rf{2.4}. To perform it, we consider separately the regimes of
small distances $r_{3+d}\ll a$ compared to the size of the internal space and of large distances $r_{3}\gg a$. For $r_{3+d}\ll a$ the field lines penetrate all $D$ space
dimensions, whereas at length scales $r_{3}\gg a$ the internal dimensions are not accessible and effectively the field lines are spreading  only along the $3$ spatial
dimensions of the external space $M_{3}$.

As in \cite{ADD}, we pass for simplicity of the calculations from the compact $d-$torus to its covering space $\mathbb{R}^{d}$. The single mass $m_1$ in $T^{d}$ is then
mapped to a mass lattice in $\mathbb{R}^d$, which is build from $m_1$ and its mirror images in the cover.

For a test mass $m_2$ at a small distances $r_{3+d}\ll a$ from $m_1$ the mirror masses give a negligible contribution to the gravitational force and in first order
approximation the Gauss law in $D$ space dimensions holds in accordance with \rf{g1}. In this case the infinite set of mirror masses can in rough approximation be
considered as symmetrically distributed around $m_1$, $m_2$ so that their contributions will compensate each other.

For the analysis of the opposite case of a large separation $r_{3}\gg a$
between the masses $m_1$ and $m_2$, we assume the mass $m_1$
placed in the origin of the $D-$dimensional space. The lattice
of the mirror masses spans then a $d-$dimensional subspace
$\mathbb{R}^d$ ("wire") in the space $\mathbb{R}^{d}\times \mathbb{R}^{3}$
and the total effective force $F_{eff}$ will result from the
gravitational attraction of the whole lattice-subspace. For $r_{3}\gg a$,
the mirror masses appear homogeneously smeared about this
subspace with a "surface" density $m_1/a^{d}$. Let us choose
coordinates
%%%%%%%%%
\be{g3} ({\bf z},{\bf r}_3) \in \mathbb{R}^{d}\times \mathbb{R}^{3} \ee
%%%%%%%%
and a test mass $m_2$ located at the point $(0,{\bf r}_3)$. The force orthogonal to the lattice produced on $m_2$ by a small volume element $dV_{d}$ at a point $({\bf
z},0)$ of the (lattice) covering space $\mathbb{R}^{d}$ is given by
%%%%%%
\ba{g4} &{}&dF(z,r_3)=G_{\mathcal{D}}\frac{r_3}{(z^2+r_3^2)^{1/2}}\frac{m_1 dV_{d}}{a^{d}}\frac{m_2}{(z^2+r_3^2)^{(2+d)/2}}\nn \\ &=&\frac{G_{\mathcal{D}}m_1 m_2}{a^{d}}
\frac{r_3 dV_{d}}{(z^2+r_3^2)^{(3+d)/2}}\, . \ea
%%%%%%
The factor $r_3/(z^2+r_3^2)^{1/2}$ is the cosine between the force direction and the lattice normal. Due to the lattice symmetry, the force component parallel to the
lattice is compensated by an opposite force component from the point $(-{\bf z},0)$. The total effective force $F_{eff}(r_3)$ of the mass lattice on the test particle
can now be easily obtained by integrating over the volume of $\mathbb{R}^{d}$. Choosing spherical coordinates on $\mathbb{R}^{d}$, the volume of a thin shell with radius
$z$ is
%%%%%%
\be{g5}
dV_{d}=S_{d}z^{d-1}dz,
\ee
%%%%%%
where $S_d$ is square of $(d-1)$-dimensional sphere of a unit radius (see \rf{2.1})
%%%%%%
%\be{g5a}
%S_{d_1}=\frac{2\pi ^{d_1/2}}{\Gamma (d_1/2)},
%\ee
%%%%%%%
and the total effective force can be calculated as
%%%%%
\be{g6}
F_{eff}(r_3)=\frac{G_{\mathcal{D}}m_1 m_2 }{a^{d}}r_3 S_{d} \int^\infty
_0\frac{z^{d-1}dz}{(z^2+r_3^2)^{(3+d)/2}} \, .
\ee
%%%%%%
With the substitution $t=z^2$ the integral can be transformed to the standard integral \cite{PBM}
%%%%%
\be{g7}
\int^\infty _0\frac{x^{\alpha -1}dx}{(x+b)^\rho}=b^{\alpha -\rho}
B(\alpha, \rho -\alpha), \qquad 0< \Re \alpha < \Re \rho
\ee
%%%%%
so that for the corresponding term in \rf{g6} we find
%%%%%%
\ba{g8} &{}&\int^\infty _0\frac{z^{d-1}dz}{(z^2+r_3^2)^{(3+d)/2}}=\frac 12 \int^\infty _0\frac{t^{d/2 -1}dt}{(t+r_3^2)^{(3+d)/2}}\nn \\ &=&\frac{1}{2r_3^{3} }
B\left(\frac{d}{2},\frac{3}{2}\right) =\frac{1}{r_3^{3}}\frac{\Gamma \left(\frac{d}{2}\right)\Gamma \left(\frac{3}{2}\right)}{2\Gamma \left(\frac{3+d}{2}\right)}\, . \ea
%%%%%%
Expressing the $\Gamma-$functions in \rf{g8} in terms of surface areas: $\Gamma (D/2)=2\pi^{D/2}/S_D$ gives
%%%%%%%
\be{g9}
\int^\infty
_0\frac{z^{d-1}dz}{(z^2+r_3^2)^{(3+d)/2}}=\frac{1}{r_3^{3}}\frac{S_{3+d}}{S_{d}S_{3}}\, .
\ee
%%%%%%
Hence, the total effective force takes the simple form
%%%%%%%
\be{g10}
F_{eff}(r_3)
=\frac{G_{\mathcal{D}}S_{3+d}}{S_{3}a^{d}}\; \frac{m_1 m_2}{r_3^{2}}
\ee
%%%%%%%
and equating it with the gravitational force law \rf{g1a} in three-dimensional space, we reproduce the result \rf{2.4}
for the relation between three-dimensional and multidimensional gravitational constants.

%%%%%%%%%%%%%%%%%%%%%%%%%%%%%%%%%%%%%%%%%%%%%%%%%%%%%%%%%%%%%%%%%%%%%%%%%%%%%%%%

\end{document}